\documentclass{webofc}

\usepackage[varg]{txfonts}   
\usepackage{hyperref}
\usepackage{url}
\hypersetup{colorlinks=true,citecolor=blue,urlcolor=blue,linkcolor=blue}
\begin{document}
\title{Theory of Quarkonia as Probes for Deconfinement}
%
%

\author{\firstname{Xiaojun} \lastname{Yao}\inst{1}\fnsep\thanks{\email{xjyao@uw.edu}} 
}

\institute{InQubator for Quantum Simulation, Department of Physics, University of Washington, Seattle, WA 98195, USA. 
}

\abstract{This is a plenary talk given at Quark Matter 2025, summarizing recent theoretical developments for the understanding of quarkonium production in relativistic heavy ion collisions and how quarkonium uniquely probes the deconfined phase of QCD matter.
}
\maketitle
\section{Introduction}
\label{intro}
Quarkonium denotes a bound state consisting of a heavy quark ($Q$) and a heavy antiquark ($\bar{Q}$). In vacuum, the mass spectra of quarkonium states below the open heavy meson threshold can be well described by nonrelativistic potential models such as the Cornell potential, which contains a Coulomb part and a confining part. In 1986, Matsui and Satz studied the plasma screening effect on quarkonium and found that at sufficiently high temperature the potential can be significantly suppressed, as a result, quarkonium states no longer exist in the hot medium, i.e., ``melt''~\cite{Matsui:1986dk}. Since then, quarkonium suppression has been used as a probe of the quark-gluon plasma (QGP) produced in relativistic heavy ion collisions.

\section{Theoretical Developments}
\label{sec:2}
\subsection{Models with Completely Melt Quarkonium}
\label{sec:complete_melt}

Let us assume for the moment that quarkonium does not exist inside the QGP and is only (re)generated during the smooth cross-over from the QGP phase to the hadron gas. This line of thinking is highlighted in the Statistical Hadronization Model (SHM)~\cite{Andronic:2017pug,Andronic:2019wva}.

\subsubsection{Statistical Hadronization Model}
\label{sec:SHM}
The SHM assumes that charm quarks are unbound inside the QGP and has reached kinetic equilibrium. Charmonium is produced via an instantaneous statistical combination of open charm quarks at freezeout. A comparison between the SHM results and the experimental data on $J/\psi$ yield in 5 TeV Pb-Pb collisions shows good agreement at low transverse momentum $p_T<5$ GeV. At higher $p_T$, the model underestimates the data, since the $p_T$ spectra of the charm quarks that combine into charmonia are assumed thermal and are too soft. For the nuclear modification factors $R_{AA}$, the model results show good agreement with the data for $J/\psi$ but underestimate for $\psi(2S)$~\cite{ALICE:2022jeh}. This is because the upper limit of the experimental $p_T$ cut is 12 GeV for $\psi(2S)$, higher than that for $J/\psi$, which is 8 GeV.

\subsubsection{Instantaneous Coalescence Model}
\label{sec:icm}
The information of the quarkonium wavefunction can be incorporated into the instantaneous hadronization by using a Wigner function, which is known as the Instantaneous Coalescence Model~\cite{Greco:2003vf}. See also the Parton-Hadron-String Dynamics model~\cite{Song:2017phm,Song:2023zma}.

The $B_c$ meson consisting of a bottom quark and an anticharm quark (or a charm and an antibottom) is similar to quarkonium. The size of the ground state of the $B_c$ meson ($\sim0.34$ fm) is closer to that of $J/\psi$ ($\sim0.4$ fm) than $\Upsilon(1S)$ ($\sim0.2$ fm). The $B_c$ meson production in 5 TeV Pb-Pb collisions was recently studied~\cite{he,Zhao:2024jkz}, where a Langevin equation is used to describe the transport of unbound heavy quarks inside the QGP so that one can use perturbative QCD calculated $p_T$ spectrum as the initial condition of the evolution, which is harder than the thermal one. A bottom quark surrounded by multiple charm quarks is instantaneously coalescenced with one charm into a $B_c$ meson at freezeout. It was found that the $B_c$ meson production is enhanced by a factor of $5$ roughly at $p_T<5$ GeV. It would be very interesting to try measuring the $B_c$ meson production experimentally.

\subsection{Lattice Studies of Screening}
\label{sec:lattice}
Now we come back to the question of whether quarkonium can exist inside the QGP or not. If we consider a deeply bound state at low temperature such that the quarkonium state can be thought of as a color dipole, the interaction strength between the quarkonium state and the QGP can be very weak. For example, we have $rT\sim0.2$ for $\Upsilon(1S)$ at $T=200$ MeV. So it would be possible for a quarkonium state to stay bound inside the QGP.

A recent lattice QCD calculation in the Coulomb gauge in the static limit (infinite heavy quark mass $M\to\infty$) showed no evidence of screening in the real part of the potential, up to $T=350$ MeV~\cite{larsen,Bazavov:2023dci}. A lattice NRQCD calculation at finite bottom quark mass showed no mass change in the bottomonium family up to $T=260$ MeV~\cite{larsen,Ding:2025fvo}. On the other hand, the lattice QCD results~\cite{larsen,Bazavov:2023dci} confirmed a nonzero imaginary part of the potential that grows with the temperature, which has been perturbatively calculated before~\cite{Laine:2006ns,Beraudo:2007ky,Brambilla:2008cx}. The imaginary part of the potential results in a thermal width for quarkonium states, which means that quarkonium, even though it can exist as a well-defined bound state inside the QGP, may dissociate into an unbound heavy quark pair via dynamical processes that transfer enough energy into quarkonium from the medium. The inverse of the dissociation process, recombination~\cite{Thews:2000rj}, may also happen if the quarkonium state can exist and dynamically radiates enough energy into the medium. Now I will discuss a few transport calculations that implement both dissociation and recombination.

\subsection{Transport Models with Dissociation and Recombination}
\label{sec:transport}
\subsubsection{Texas A$\&$M University Model}
\label{sec:tamu}
The Texas A$\&$M University (TAMU) model~\cite{He:2021zej} implements dissociation and recombination through a relaxation rate equation that evolves the quarkonium number towards the equilibrium value, which can be time-dependent. The dissociation rate is calculated from the $2\to2$ process $g+\mathcal{B}\to Q+\bar{Q}$, where $\mathcal{B}$ denotes a quarkonium state, as well as the $2\to3$ processes in the quasi-free limit that treats the quarkonium state as two independent heavy quarks, i.e., $2\times(q/g+Q \to q/g +Q)$. The quasi-free rate is multiplied by a phenomenological factor to account for the interference effect between the $Q\bar{Q}$ pair. In the most recent developments, the $T$-matrix summation formalism was used, which can take lattice inputs to fix parameters~\cite{Wu:2025hlf}. The model calculations can well describe the experimental data on $R_{AA}$ and the elliptic flow coefficient $v_2$ for $p_T$ below roughly three times the quarkonium state mass, see e.g., Refs.~\cite{ALICE:2020pvw,ALICE:2022jeh}.

\subsubsection{Santiago Comover Interaction Model}
\label{sec:comover}
The Santiago Comover Interaction model implements a similar rate equation~\cite{Ferreiro:2018wbd}. Instead of tracking the quarkonium state number, the rate equation tracks the density. The input is the comover-quarkonium interaction cross section. In practice, no recombination is implemented for bottomonium. The model calculations can well describe $R_{AA}$ and $R_{pA}$ for the $\Upsilon(nS)$ states.

\subsubsection{Tsinghua Model}
\label{sec:tsinghua}
The Tsinghua model~\cite{Liu:2010ej} implements dissociation and recombination in a relativistic Boltzmann equation. Only the $2\leftrightarrow 2$ process $g+\mathcal{B} \leftrightarrow Q+\bar{Q}$ is included. The model calculation can well describe the $R_{AA}$ of $J/\psi$ measured in 200 GeV Au-Au collisions~\cite{STAR:2019fge}.

\subsection{Open Quantum System Framework}
\label{sec:oqs}
With these successful phenomenological calculations for quarkonium production in heavy ion collisions, we ask what properties of the QGP are probed. Some may think it is the temperature profile. But in most calculations, temperature is used as an input, obtained from hydrodynamics. Some may think it is the dissociation/recombiantion rate. But these rates are properties of quarkonia, rather than those of the QGP. Thus, to answer the question, it is important to separate the intrinsic properties of quarkonium from the intrinsic properties of the QGP. This can be achieved by using the open quantum system (OQS) framework and effective field theory (EFT), which I will explain now.

The OQS framework is widely applied in quantum optics and quantum information science. It was first applied to quarkonium production in heavy ion collisions by Akamatsu and Rothkopf~\cite{Akamatsu:2011se,Akamatsu:2014qsa}. Later, it was combined with EFT, which significantly simplifies the calculations in a systematic way~\cite{Brambilla:2016wgg,Brambilla:2017zei,Blaizot:2017ypk,Miura:2019ssi,Yao:2018nmy}. Recent reviews can be found in Refs.~\cite{Rothkopf:2019ipj,Akamatsu:2020ypb,Sharma:2021vvu,Yao:2021lus}.
The OQS framework treats the heavy quark pair, bound or unbound, and the QGP as a closed quantum system described by the density matrix $\rho_{\rm tot}$. By evolving the total state in time and tracing out the QGP degrees of freedom, one obtains the evolution equation for the $Q\bar{Q}$ pair
\begin{align}
\rho_{Q\bar{Q}}(t) = {\rm Tr}_{\rm QGP}[ U(t) \rho_{\rm tot}(0) U^\dagger(t) ] \,,
\end{align}
where $U(t)$ denotes the unitary time evolution of the total system. Then one can make various approximations to simplify the description of $\rho_{Q\bar{Q}}(t)$, where EFT plays a major role.

EFT is constructed from a hierarchy of energy scales. For quarkonium in vacuum, we have three relevant scales: heavy quark mass $M$, inverse of quarkonium typical size $r^{-1}$ and the typical binding energy $E_b>0$. They are widely separated $M\gg r^{-1} \gg E_b$. In a hot medium, we have additional thermal scales such as the temperature $T$ and Debye mass $m_D$ for a weakly coupled plasma. Nonperturbatively the scale is $T$. We will use $T$ to represent all thermal scales in the following. Depending on where $T$ fits, one has different effective descriptions for quarkonium in-medium dynamics. When the temperature is much higher than $r^{-1}$, the resolving power of the QGP is strong enough to see the internal structure of quarkonium. As a result, each heavy quark inside the quarkonium state interacts with the QGP almost independently. The effective dynamics is described as random distortion of the wavefunction that leads to decoherence. Due to the similarity with the classical Brownian motion, this limit is called the quantum Brownian motion limit. On the other hand, when the temperature is much smaller than $r^{-1}$, the resolving power of the QGP is weak and the $Q\bar{Q}$ pair interacts with the QGP as a whole. The effective dynamics is described as transitions between different states in the $Q\bar{Q}$ level spectrum. This limit is called the quantum optical limit. Different energy hierarchies and the corresponding EFTs that have been studied are summarized in Table.~1 of Ref.~\cite{Yao:2025jyx}, together with the resulting classical and quantum evolution equations. Here we discuss a few examples. More details can be found in Ref.~\cite{Yao:2025jyx}.

\subsubsection{Osaka and Nantes Approaches}
\label{sec:lindblad_nrqcd}
The Osaka~\cite{Miura:2022arv} and Nantes~\cite{Delorme:2024rdo} groups considered the hierarchy $M\gg T \gg E_b,\Lambda_{\rm QCD}$ and numerically studied a one-dimensional (1D) Lindblad equation in the quantum Brownian motion limit. Both studies solved the Lindblad equation to late time and observed the steady state. The Osaka group used momentum basis and found the late-time occupancy of quarkonium states agrees with the Boltzmann distribution for different initial conditions and temperatures. The Nantes group used position basis and observed that at late time the diagonal part of the density matrix becomes flat at large distance, while at small distance the color singlet channel has a pronounced peak and the octet channel has a valley. These can be understood from the attractive (repulsive) potential in the color singlet (octet) channel. 

\subsubsection{Munich-Kent Approach}
\label{sec:lindblad_pnrqcd}
The Munich-Kent collaboration considered the hierarchy $M\gg r^{-1} \gg T,\Lambda_{\rm QCD}$ with $T\gg E_b$ and studied a 3D Lindblad equation in the quantum Brownian motion limit, where the quarkonium-QGP interaction is described by a color dipole. The Lindblad equation depends on two nonperturbative parameters $\kappa_{\rm adj}$ and $\gamma_{\rm adj}$ defined in terms of chromoelectric correlators to be introduced later. The scale $r^{-1}$ is assumed perturbative. In a recent work~\cite{Brambilla:2024tqg}, the three-loop QCD potential was used, which gives a more accurate description of bottomonium spectrum than the previously used Coulomb potential. For certain values of $\kappa_{\rm adj}$ and $\gamma_{\rm adj}$, the approach can well describe the $R_{AA}$ of $\Upsilon(nS)$ measured at the LHC energies. Thermalization in this framework has also been studied recently~\cite{Brambilla:2025sis}. A new development that perturbatively goes beyond the dipole interaction was discussed in this conference~\cite{jorge}.

\subsubsection{Duke-MIT Approach}
\label{sec:boltzmann_pnrqcd}
The Duke-MIT approach considered the hierarchy $M\gg r^{-1} \gg T, \Lambda_{\rm QCD}$ and the quantum optical limit. Using the OQS framework, a Boltzmann equation can be derived from first principles in the semiclassical limit, for which quantum corrections can also be obtained~\cite{Yao:2020eqy}. The derivation shows that both the dissociation rate and the recombination contribution can be written in a factorized way: They contain a piece that only depends on the wavefunctions of the bound and unbound $Q\bar{Q}$ pair and a piece that only depends on the medium properties, which is the chromoelectric correlator to be defined later. This generalizes the perturbative calculations mentioned above for the $2\leftrightarrow 2$ and $2\leftrightarrow 3$ processes to all orders and the chromoelectric correlator can be interpreted as a generalized gluon distribution, see e.g., Fig.~1 of Ref.~\cite{Nijs:2023dbc}.

The Boltzmann equation for quarkonium has been coupled with that for open heavy quarks to study quarkonium thermalization~\cite{Yao:2017fuc} and production in heavy ion collisions~\cite{Yao:2020xzw}. The calculation results for the $R_{AA}$ of $\Upsilon(nS)$ agree well with experimental data.

\subsection{Chromoelectric Correlator for Quarkonium}
\label{sec:EE}
The chromoelectric correlator for quarkonium in-medium dynamics mentioned above is defined as, e.g., (see Ref.~\cite{Scheihing-Hitschfeld:2023tuz} for more details in the notation)
\begin{align}
    [g_{\rm adj}^{++}]^>(t) &\equiv \frac{g^2 T_F }{3 N_c}  \big\langle E_i^a(t)W^{ab}(t,0)  E_i^b(0) \big\rangle_T \,,
\end{align}
where $E_i^a$ denotes the chromoelectric field with spatial index $i$ and adjoint color index $a$ and $W^{ab}(t,0)$ is an adjoint Wilson line along the time direction. The chromoelectric correlator for quarkonium is defined in a nonperturbative and gauge-invariant way. It is similar to the correlator for heavy quark diffusion, but differs crucially in terms of the operator orderings~\cite{Eller:2019spw,Scheihing-Hitschfeld:2022xqx}. The correlator for quarkonium has been calculated perturbatively at next-to-leading order~\cite{Binder:2021otw} and nonperturbatively via AdS/CFT~\cite{Nijs:2023dks,Nijs:2023dbc}. It represents a property of the QGP that can be uniquely probed by studying quarkonium production at low transverse momentum.

\subsection{From AA to pp Collisions}
Recent measurements in proton-proton (pp) collisions hinted at an event activity dependence in quarkonium production. If this is confirmed, it will indicate the breakdown of factorization that has been widely applied in calculations of quarkonium production in pp collisions. A new framework to study quarkonium production in pp collisions with environment effects was discussed in this conference~\cite{zhao}.

\section{Summary and Conclusions}
In this talk, I summarized recent theoretical progress in understanding quarkonium as a probe of deconfinement in heavy ion collisions, and highlighted a property of the QGP that can be uniquely probed by quarkonium in the dipole limit: the chromoelectric correlator. Details of each model can be found in Ref.~\cite{Andronic:2024oxz}, which summarizes the efforts of the EMMI Rapid Reaction Task Force that started at GSI in 2019.\\

This work is supported by the U.S. Department of Energy, Office of Science, Office of Nuclear Physics, InQubator for Quantum Simulation (IQuS) (https://iqus.uw.edu) under Award Number DOE (NP) Award DE-SC0020970 via the program on Quantum Horizons: QIS Research and Innovation for Nuclear Science.

%
%

\end{document}